\documentclass[graybox]{svmult}

\usepackage{mathptmx}       
\usepackage{helvet}         
\usepackage{courier}        
\usepackage{makeidx}         
\usepackage{graphicx}        
\usepackage{multicol}        
\usepackage[bottom]{footmisc}
\usepackage{url}

\usepackage{amsfonts}

\makeindex             

\def\F{\mathbb{F}}
\def\Z{\mathbb{Z}}
\begin{document}

\title*{A Digital Signature Scheme based on two hard problems}

\author{Dimitrios Poulakis and Robert Rolland}
\institute{Dimitrios Poulakis \at Department of Mathematics,
Aristotle University of Thessaloniki,
54124 Thessaloniki, Greece.
\email{poulakis@math.auth.gr} \and Robert Rolland \at Universit{\'e} d'Aix-Marseille, Institut de Math\'{e}matiques
de Marseille, case 907, F13288 Marseille cedex~9, France. \email{robert.rolland@acrypta.fr}}

\maketitle

\renewcommand{\theequation}{\arabic{equation}}
\setcounter{equation}{0}

\abstract{In this paper we propose a signature scheme based on two intractable problems,
namely the integer factorization  problem and the discrete logarithm problem for elliptic curves. 
It is suitable for applications requiring long-term security and provides  smaller signatures than the existing schemes based
on the integer factorization and integer discrete logarithm problems.}

\section{Introduction}

Many applications of the Information Technology, such as  encryption of sensitive medical data or digital signatures
for contracts, need long term cryptographic security. Unfortunately, today's cryptography provides strong tools only for short term security 
\cite{Bu}. Especially, digital signatures do not guarantee the desired long-term
security. In order to achieve this goal Maseberg \cite{Mas} suggested the use of more than one sufficiently independent signature schemes.
 Thus, if one of them is broken, then it can be replaced by a new secure one.  Afterward the
document has to be re-signed. Again we have more than one valid signatures of our document.
Of course, a drawback of the method is that the document has to be re-signed. 

In order to avoid this problem, it may be interesting for applications with long-term, to base 
the security of cryptographic primitives on two difficult problems, so if any of these problems is broken, 
the other will still be valid and hence the signature will be protected. 
We propose in this paper an efficient  signature scheme built taking into account this constraint.
The following signature scheme is based on the integer factorization  problem and
the discrete logarithm problem on a supersingular elliptic curve.
Remark that these two problems
have similar resistance to attack, thus they can coexist within the same protocol.
The use of a supersingular curve allows us to easily build a pairing that we use to verify the signature.

Several signature schemes combining the intractability of the integer 
factorization problem and integer discrete logarithm problem
 were proposed but they have proved either to be enough to solve the one of two problems for breaking the system or to have 
 other security problems \cite{Chen, Harn, Lee, Lee2, Li, Madhur, Shao,  Wei}. An interesting  scheme based on the above problems is GPS \cite{Girault}.
Furthermore, some recent such schemes are given in 
\cite{Ismail, Ismail1, Madhur, Vishnoi, Verma, Wei}.

In section \ref{sscheme} we describe the infrastructure for the implementation of the scheme.
Then we present the key generation, the generation of a signature and the verification.
In section \ref{pairing}
we show how to build a elliptic curve adapted to the situation and how to define
a valuable pairing on it. In section \ref{maptopoint} we address the problem of the map to point function
and give a practical solution. We deal with the performance of our scheme and compare it
with others in section \ref{performance}. 
In section \ref{example} we give a complete example that shows 
that the establishment of such a system can be made in practice. In section \ref{secu} we study the security of the scheme. 
Finally section \ref{conclusion} concludes the paper.

\section{The Proposed Signature Scheme}\label{sscheme}
In this section we present our signature scheme. 
 \subsection{Public and private key generation} 
A user $\mathcal{A}$, who wants to create  a public and a private key selects:

\begin{enumerate}

\item primes $p_1$ and $p_2$ such that the factorization of $n = p_1p_2$ is  unfeasible; 

\item an elliptic curve  $E$ over a finite field $\F_q$, a point  
  $P\in E(\F_q)$ with ${\rm ord}(P) = n$  
and an efficiently computable pairing $e_n$ such that $e_n(P,P)$ is a primitive $n$-th root of  1; 

\item  $g\in \{1,\ldots,n-1\}$ with $\gcd(g,n) = 1$ , $a\in \{1,\ldots,\phi(n)-1\}$ and computes $Q = g^aP$;

\item two one-way, collision-free hash functions,  $h :  \{0,1\}^* \rightarrow \{0,\ldots,n-1\}$ and  $H : \{0,1\}^* \rightarrow <P>,$ where
$<P>$  is the subgroup of $E(\F_q)$ generated by $P$. 
\end{enumerate} 
$\mathcal{A}$ publishes the elliptic curve $E$, the pairing $e_n$ and the hash functions $h$ and $H$.
The public key of $\mathcal{A}$ is $(P,Q,g,n)$ and his private key $(a,p_1,p_2)$.

\subsection{Signature generation} The user $\mathcal{A}$ wants to sign a  message
$m\in \{0,1\}^*$.   Then he chooses at random $k,l\in \{1,\ldots,\phi(n)-1\}$ such that
 $k+l =  a$. Next, he computes 
\begin{center}
$s =  k+h(m)+n \ {\rm mod}\ \phi(n)$ \ \ \  and  \ \ \ $S = g^l H(m)$.
\end{center}
Let $x(S)$ be the $x$-coordinate of $S$ and $b$ a bit determining $S$. The signature of $m$ is  $(s,x(S),b)$.

\subsection{Verification} 
Suppose that $(s,x,b)$ is the signature of $m$. The receiver uses $b$ in order to determine $y$ such that $S = (x,y)$ is a point of $E(\F_q)$. He
accepts the signature if  and only if
$$e_n(g^sP, S) = e_n(g^{h(m)+n}Q,H(m)).$$

\ \\
{\it Proof of correctness of verification.}  
Suppose that the signature $(x,s,b)$ is valid and $S =(x,y)$ is a point of $E(\F_q)$. Then  we get 
$$e_n(g^sP,S)  = e_n(g^{k+h(m)+n} P,g^lH(m)) =  e_n( g^{h(m)+n}Q,H(m)).$$

Suppose now we have a couple $(s,S)$, where $s\in \{1,\ldots,\phi(n)\}$  and $S \in <P>$, such that 
$$e_n(g^sP,S)  =  e_n(g^{h(m)+n}Q,H(m)).$$
Since $H(m),S \in <P>$, there are $u,v\in \{0,\ldots,n-1\}$ such that 
$S = uP$ and $H(m) = vP$. Thus we get
 $$e_n((g^su-g^{h(m)+n+a}v)P,P) = 1.$$
 The element $e_n(P,P)$ is a primitive $n$-th root of  1 and so, we obtain
$$ uv^{-1} \equiv g^{a+h(m)+n-s}\ ({\rm mod} \ n),$$
Putting $l= a+h(m)+n-s \ {\rm mod}\ \phi(n)$ and  $k = a-l \ {\rm mod}\ \phi(n) $,  we get
\begin{center}
$s =  k+h(m)+n \ {\rm mod}\ \phi(n)$ \ \ \  and  \ \ \ $S = g^l H(m)$.
\end{center}
 It follows  that $(s,x(S),b)$ is the signature of $m$ (where $b$ is a bit determining $S$).

\section{The elliptic curve and the pairing}\label{pairing}

In this section we show how we can construct an elliptic with the desired properties
in order to implement our signature scheme. This task is achieved by the following algorithm:
\begin{enumerate}
 \item select two large prime numbers $p_1$ and $p_2$
such that the factorization of $p_1-1$, $p_2-1$ are known and the computation of the 
factorization of $n = p_1p_2$ is unfeasible;
 \item select a random prime number $p$ and compute $m = $ ord$_n(p)$;
 \item  find, using the  algorithm of \cite{Broker}, a supersingular elliptic curve $E$ over
$\F_{p^{2m}}$ with trace $t=2p^m$;
 \item return $\F_{p^{2m}}$ and $E$.
\end{enumerate}

Since  the trace of $E$ is $t=2p^m$, we get  $|E(\F_{p^{2m}})| =(p^m-1)^2$. On the other hand, we have 
$m = $ ord$_n(p)$, whence  $n|p^m-1$, and so $n$ is a divisor of $|E(\F_{p^{2m}})|$. Therefore
 $E(\F_{p^{2m}})$ contains a subgroup of order $n$. 

By \cite[Theorem 1.1]{Broker}, we obtain, under the assumption that
the Generalized Riemman Hypothesis is true, that the time complexity of Step 3 is $\tilde{O}((\log p^{2m})^3)$.
Furthermore, since the factorization of $\phi(n) = (p_1-1)(p_2-1)$ is known, the time needed for the
computation of $m$ is $O((\log n)^2/\log\log n)$ \cite[Section 4.4]{kara}.

For the implementation of our signature scheme we also need a point $P$ with order $n$ and an
efficiently computable pairing $e_n$ such that $e_n(P,P)$ is a primitive $n$-th root of  1.
 The Weil pairing does not fulfill this requirement and also, in many instances, the Tate pairing; the same happens
for the  eta pairing (the ate and omega pairings can be computed only on the ordinary elliptic curves) 
\cite{Bar,  Hess, Zhao}. Let $\epsilon_n$ be one
of the previous pairings on $E[n]$. Following the method introduced by E. Verheul
 \cite{ver01},  we use a distortion map $\phi$ such that the points $P$ and $\phi(P)$ is a generating set for $E[n]$
and we consider the pairing 
$e_n(P,Q)=\epsilon_n(P,\phi(Q)).$
The algorithm of \cite[Section 6]{Galb} provides us a method for the determination of $P$ and $\phi$.

Another method for the construction of the elliptic curve $E$ which is quite efficient in practice is given
by the following algorithm:
\begin{enumerate}
 \item draw at random a prime number $p_1$ of a given size $l$ (for example $l$
is $1024$ bits);
 \item draw at random a number $p_2$ of size $l$;
 \item repeat $p_2={\rm NextPrime}(p_2)$ until $4p_1p_2-1$ is prime;
 \item return $p=4p_1p_2-1$.
\end{enumerate}

It is not proved that this algorithm will stop with a large probability. This is an open problem
which is for $p_1=2$ the Sophie Germain number problem. But in practice
we obtain a result $p$ which is a prime of length $2l$. 

Since $p \equiv  3\  ({\rm mod}\ 4)$,  the elliptic curve defined over $\F_p$ by the equation 
$$y^2=x^3+ax,$$  where $-a$ is not a square in $\F_p$, is
 supersingular  with $p+1=4p_1p_2$ points.
By \cite[Theorem 2.1]{vla99_1}, the group $E(\F_p)$ is either cyclic
or $E(\F_p)\simeq \Z/2p_1p_2\Z \times \Z/2\Z$. In each case the group $E(\F_p)$
has only one subgroup of order $n=p_1p_2$, and this subgroup is cyclic.

If $\epsilon_n$ is one of the Weil, Tate or eta pairings on $E[n]$, then we use the distortion map 
$\phi(Q)=\phi(x,y)=(-x,iy)$ with $i^2=-1$ (cf. \cite{jou02}) and so, we obtain the following pairing:
$e_n(P,Q)=\epsilon_n(P,\phi(Q)).$

\section{The map to point function}\label{maptopoint}
Let $G$ be the subgroup of order $n=p_1p_2$ of $E(\F_q)$ introduced in the previous section.
In order to sign using the discrete logarithm problem on this group,
we have to define a hash function into the group $G$, namely a map to point function.
This problem was studied by various authors giving their own method, for example
in \cite{bls01} or  \cite{ica09}. We give here the following solution. Let us denote by
$|n|=\lfloor \log_2(n) \rfloor+1$ the size of $n$.  Let $h$ be a key derivation function,
possibly built using a standard hash function. We recall that $h$ maps
a message $M$ and a bitlength $l$ to a bit string $h(M,l)$ of length $l$. Moreover we will suppose
that $h$ acts as a good pseudo-random generator.
Let $Q$ be a generator of the group $G$.
Let us denote by $(T_i)_{i \geq 0}$ the sequence of bit strings defined by
$T_0=0$ and for $i\geq 1$ 
$$T_i=a_u\cdots a_0,$$
where $i=\sum_{j=0}^u a_j2^j$ and $a_u=1$.

To map the message $m$ to a point $H(m)$ we run the following algorithm:

\medskip

\noindent $\quad$ $i:=0$;\\
$\quad$ Repeat\\
$\quad\quad$ $k:=h(m||T_i,|n|)$;\\
$\quad\quad$ $i:=i+1$;\\
$\quad$ Until $k < n$;\\
$\quad$ Output $H(M)=k.Q$;

\medskip

This Las Vegas algorithm has a probability zero
to never stop. In practice this algorithm stops quickly, namely as
$2^{|n|-1} < n < 2^{|n|}$ then
the expected value of the number of iterations 
is $<2$. If one can find a collision for $H$
it is easy to find a collision for $h$.

\section{Performance Analysis}\label{performance}

In this section we analyze the performance of our scheme.
The computation of $s$  requires two additions modulo $\phi(n)$. 
The computation of $S$ needs a modular  exponentiation $g^l\ ({\rm mod}\ n)$ and 
the computations of  $H(m)$ and  $g^l H(m)$.  Note that the computation of
$g^l\ {\rm mod}\ n$ and $k+n \ {\rm mod}\ \phi(n)$ can be done off-line. 
Thus, the signature generation
requires only a modular addition and a point multiplication on the elliptic curve. The signature verification needs two modular  exponentiations, 
two points multiplications on the elliptic curves, and two pairing computations. Moreover note that the length of the  signature
 of a message is the double of its length.

The signature generation in the GPS scheme \cite{Girault} needs only one modular exponentiation and the signature verification two. The signature length
is the triple of the message length.
The most efficient of the schemes given in \cite{Ismail, Ismail1, Madhur, Vishnoi, Verma, Wei} requires 3 modular exponentiations for the signature generation
and 4 modular exponentiations for the signature verification. The signature length of the above schemes is larger than the double of the message length.  

Hence we see that the signature length in our scheme is smaller than that in GPS and the other schemes.
 Moreover, the performance of the proposed algorithm is  competitive to the
performance of the above schemes.

\section{Example}\label{example}

In this section we give an example of our signature scheme. We consider the 1024-bits
primes
$$
\begin{array}{lll}
p_1 &:= & 61087960575038789816988536114150792266377636351843177587564\\
& & 31924627119957041754060999158399749767833896533906296859311\\
& & 25485163415231551275212583044052150577614828617005803730389\\
& & 43877400689242960278845109703690843026188873847913442234432\\
& & 36591255684234493362159572100747699404245339214008078743836\\
& & 7162669180839
\end{array}
$$
and
$$
\begin{array}{lll}
p_2 &:= & 950794575789036193985289494100238271764913649341936446441081\\
& & 377072500578035754538268902518142982960234055319718348171564\\
& & 531835348013169675598575434394528269729126327128190711758193\\
& & 487088395696503090307111303433870155114599617217105648040005\\
& & 344506796898422897977489196110610260665664553656001074068087\\
& & 13249343.
\end{array}
$$

We take $n=p_1p_2$. The number $q=4n -1$ is a prime. 
Since $q \equiv 3 \ ({\rm mod}\  4)$, the elliptic curve $E$ defined by the equation 
$y^2=x^3+x$ over $\F_q$ is superesingular.
The point $P = (x(P),y(P))$, where $x(P) = 2^{1500}+2$ and
$$y(P) = 92629334720096485394250229023531473128561210303747369871170$$
$$532503591346084781038053790347765721405539373837575715741111302632$$
$$222520728502603977901582753916707479492439228918725855423715991340$$
$$003621514555505206507732534242013847767107764800751435936328543137$$
$$789247911179152023276247696951339536945505339588067200491193957998$$
$$044975563046555194785086909103272771864842171753848435480722850484$$
$$547366650914307823107502201128733622163636510656608071825566283432$$
$$994640380462713709910638633429178083083878848700277309884412794341$$
$$026781057881112432733889255328105052291841518470922081921433382412$$
$$472012678120546125640726148962.$$
has order $n$. We take $g = 2$, 
$$a= 2^{256}+2^9+1 = 11579208923731619542357098500868790785326998466$$
$$5640564039457584007913129640449$$
and we compute
$$g^a \ {\rm mod}\  n = 291246612437704212466554616370488460582482345$$
$$412043139387071627568366461190658309237330580043030838224854789252$$
$$968050905018578440545530480131761225347896913705349073419345335895$$
$$868832920014327349522957752032149784650672578527400186028060209053$$
$$035728070430079944852013985987562947197675511448867860271390438151$$
$$997510376157277527652722786834963496843487625119512000324307142997$$
$$876216044005309541179123902262183075125684914484636806915549910481$$
$$194533920018176890664864601123368083711476432553316859751469426810$$
$$204461407620204756483516542976417259702626996120442929825569733396$$
$$7126221051950952443115939209262561714767443.$$
Next, we compute  $Q = g^aP = (x(Q),y(Q))$, where 
$$x(Q)= 492906626963089094011867684016548035835802792163377707597056$$
$$795455537761970341320418289803336076175870732053896841006011789243$$
$$411173491601076264818884432777686675649566399360544060115589059409$$
$$495626348669253033853643920668587107209662122339196308521380419432$$
$$395876777001037759129809826188826444792896302483531297500328577661$$
$$115644137663377694781584798800831919655207788055426633821916253648$$
$$545542264181819923868715936604077661019515870909292645145292612582$$
$$082056454491673626406957411250447615805464800603537427266421084067$$
$$068889942487927367826706242600925470755091415792336658258887358233$$
$$6648011173165127581579893233$$
and
$$y(Q)= 925164000667984941436213463843562867132842692526639503713623$$
$$100761058759325653912386860742637828197211675023371765292190166225$$
$$688907658763278636042952123928199605188431021730950523522172176061$$
$$249916336352942245517540928470987327163690899169971423566730046146$$
$$040131461711982514952573761305725771859092373093590718229549775728$$
$$318091393459721685022050067573052541368464407556329663187692087325$$
$$785318806656273634451502898900933909082715458588013832847281982918$$
$$045250406217417892195982283414569723280463029281881025844011710313$$
$$003637423244716948430928877376648184124169704330493421073010959904$$
$$2000468957343998962535886947.$$
Therefore  $(P,Q,2,n)$ and  $(a,p_1,p_2)$ are a public key and the 
corresponding private key for our signature scheme. Moreover, we
can use the Tate pairing with the distortion map $\phi(x,y)=(-x,iy)$ with $i^2=-1$.

\section{Security of the Scheme}\label{secu}

 In this section we shall discuss the security of our system. First, 
we remark that if an attacker wants to compute the private key $(a,p_1,p_2)$  from the public key, he has to factorize $n$ and
to compute  the discrete logarithm $g^a$ of $Q$ to the base $P$ and next to calculate the discrete logarithm $a$ of $g^a$
to the base $g$ in the group $\Z_n$. Note that  an algorithm which computes the discrete logarithm modulo $n$ 
implies an algorithm which breaks the Composite Diffie-Hellman key distribution scheme for $n$ and any  algorithm which 
break his scheme for a non negligible proportion of the possible inputs can be used to factorize $n$ \cite{McCurley, 
Biham}. 

In order to study the security of the scheme we are going to 
look at the two worst cases:
\begin{enumerate}
 \item the factorization problem is broken but the elliptic curve discrete logarithm problem is not;
 \item the elliptic curve discrete logarithm problem is broken but the factorization problem is not.
\end{enumerate}

In each case we will prove that if an attacker is able to generate a valid signature for any given message $m$,
then it is able to solve, in the first case the elliptic curve discrete logarithm problem and in the second case
the factorization problem.

\medskip
 
1) Let us suppose that the attacker is able to factorize $n$. Then he can compute $\phi(n)$. But he is unable to compute $a$
since $a$ is protected by the elliptic curve discrete logarithm problem and by the discrete logarithm problem modulo $n$, because
the only known relation involving $a$ is $Q=g^aP$. So, in order to produce a valid signature of a message $m$ the attacker
has only two possibilities: he can arbitrary choose $k$, and then he can compute $s$ but not $S$, or
choose arbitrary $l$ and the he can compute $S$ but not $s$.

\medskip

2) Let us suppose now that the attacker is able to solve the elliptic curve discrete logarithm problem. Then he can compute $g^a$ 
but as the factorization problem is not broken the discrete logarithm problem modulo $n$ is not broken and consequently
he cannot compute $a$ (cf. the beginning of this section). Then as in 1) he cannot compute simultaneously $s$ and $S$.

\section{Conclusion}\label{conclusion}
In this paper we defined a signature system based on two difficult arithmetic problems.
In the framework chosen, these problems have similar resistance to known attacks.
We explained how to implement in practice all the basic functions we need for the establishment 
and operation of this system. This strategy has an interest in any application that includes a 
signature to be valid for long. Indeed, it is hoped that if any of the underlying problems is broken, 
the other will still be valid. In this case, the signature should be regenerated with a new system, 
without the chain of valid signatures being broken.  Finally, the signature length of our scheme is
smaller than that of the  schemes based on integer factorization and integer discrete logarithm problems,
and its performance is competitive to that of these schemes.

\end{document}